\documentclass[11pt]{amsart}
\usepackage{geometry}                
\usepackage{graphicx}
\usepackage{amssymb}
\usepackage{epstopdf}
\newcommand{\C}{\mathbb{C}}
\newcommand{\R}{\mathbb{R}}
\newcommand{\cC}{\mathcal{C}}
\newcommand{\eps}{\varepsilon}

\title{Shooting function for 1D Schr\"odinger operators}
\author{R.S.MacKay}
\address{Mathematics Institute, University of Warwick, Coventry CV4 7AL, UK}
\email{R.S.MacKay@warwick.ac.uk}
\date{\today}                                           

\begin{document}
\begin{abstract}
For Schr\"odinger operators with suitable 1D potentials, focussing particularly on those that go to infinity at infinity, a characteristic function is constructed, via shooting functions.  It is proved to be entire and its zeroes to be the eigenvalues.
\end{abstract}
\maketitle

\section{Preface}
Michael Berry has been and continues to be an inspiration to many.  I think my first encounter with him was reading his pedagogical review ``Regular and irregular motion" \cite{B1} as a beginning graduate student.  Then I was fortunate to attend the 1981 Les Houches school on ``Chaotic behaviour of deterministic systems'' in which he gave a course on ``Semi-classical mechanics of regular and irregular motion'' \cite{B2}.  A few years later we happened to be discussing corrections to WKB theory.  When I mentioned an asymptotic result on small reflections I had obtained in 1978 during my ``Part III'' at Cambridge (an optional fourth year for the Mathematics undergraduate course) Michael told me (nicely) that it was wrong because the later terms in my expansion were of the same order as the first and they conspire to multiply the first by $\pi/2$!  He pointed me to a correct treatment \cite{PK} (there is also a paper by Michael on the topic \cite{B3}).  So I learnt that asymptotics is more subtle than I'd realised and I am grateful for that.

Thus it is with some trepidation that I have chosen the topic of ``shooting functions for 1D Schr\"odinger operators", which potentially overlaps with a lot of Michael's areas of expertise in semi-classical mechanics.  I hope, however, that what I say will be both correct and new (and interesting!).  I hope also that the terminology of ``shooting function'' will not raise concerns from Michael, who is renowned for his frequent inclusion in Acknowledgements of ``This work was not supported by any military agency".

\section{Introduction}
The principal subject of the paper is one-dimensional Schr\"odinger operators 
\begin{equation}
L \psi = -\psi'' + V\psi
\end{equation}
for real-valued potentials $V$ on an interval or the whole of the real line $\R$, with suitable boundary conditions on $\psi$ ($' $ denotes derivative).
Mathematically, this takes us into the territory of unbounded operators and their many subtleties, as described in texts like \cite{DS2, RS1}, yet more cause for trepidation.  
Particular attention is paid to the case of the half-line with $V(x) \to +\infty$ as $x \to +\infty$ and a hard wall at $x=0$, but variants are considered.

The main point is to make concrete the concept of ``characteristic function'' $P_L$ of such an operator.  It generalises the characteristic polynomial of a finite-dimensional linear operator, in the sense that it is an entire function on the complex plane and its zeroes are the eigenvalues of the operator (with multiplicity).  Various approaches to do this exist in the literature under names like ``spectral determinant'' and ``functional determinant'' and involve techniques like ``zeta-regularisation'' \cite{W} or ``functional integration'' \cite{D, KK03}, but they seem to me excessively mysterious, at least for this context.

A definition and properties of $P_L$ will be obtained by mathematical examination of the ``shooting method'' for determining eigenvalues of such an operator (e.g.~\S 18.1 of \cite{P+}).  
In the case of regular boundary conditions, e.g.~$\psi(a)=\psi(b)=0$, this consists of evaluating the solution $\psi_E$ to $L\psi=E\psi$ for $E\in \C$ with final condition $\psi(b)=0, \psi'(b)=1$, and letting $P_L(E) = \psi_E(a)$.  Then the eigenvalues of $L$ are the zeroes of $P_L$, so $P_L$ can be defined to be the characteristic function of $L$.  This approach to defining a characteristic function of a Schr\"odinger operator is attributed by \cite{D} to Gelfand and Yaglom \cite{GY}, though I couldn't find it there.
In the case of one or both boundary conditions being singular, e.g.~$\psi(x) \to 0$ as $x \to +\infty$, one has to work out how to start in the decaying subspace or how to measure to what extent a solution from the other boundary is not in the decaying subspace.  This is the principal matter to be addressed here.

There is a large literature on this sort of issue, involving concepts such as Weyl-Titchmarsh functions, Jost functions and Evans functions (e.g.~\cite{LS}, which talks about the relations between all three and gives a guide).  In particular, \cite{D} contains a pedagogical exposition of the Gelfand-Yaglom method for the case of regular boundary conditions.  But what I have read seems to me to skate over some non-trivial issues that arise for potentials going to infinity at infinity that I address here.

The main consideration is that to compare solutions for different values of $E$ one has to fix a normalisation of the decaying solution, independent of $E$.  Under an additional condition on $V$ it will be shown that the ratio of two decaying solutions for different values of $E$ goes to a limit at infinity, so the solutions $\psi_E$ can be normalised to be asymptotic to one another at infinity.  Then if the first boundary condition is regular, $P_L(E)$ can be defined to be $\psi_E(a)$ again.  This fixes it up to an overall nonzero constant, which plays no role.  Extension to singular boundary conditions at both ends will be treated also.

Under our conditions, it will be proved that $P_L$ is entire (with a sketch that it is of order in $[\tfrac12,1]$) and its zeroes are the eigenvalues of $L$.

Some examples of potentials $V$ will be given for which the resulting characteristic functions can be written in terms of standard functions, such as Bessel functions or their generalisation to Whittaker functions, which have an extra ``magnetic field'' parameter $\kappa$ (the Bessel functions being (up to scale) the case $\kappa=0$).  The energy appears as the order of the function; its usual argument is fixed.

This introduction ends with some comments on Riemann's hypothesis, a subject of particular interest to Berry.
The proximity of the Bessel case to Riemann's $\xi$-function (evaluated at $2\sqrt{E}$) was found by Polya \cite{P}, though without noting the connection of the former to a spectral problem (strange for the claimed originator of the spectral hypothesis for Riemann's $\xi$ \cite{Po2}).  A better fit can be obtained by extending to Whittaker functions \cite{L}.
It is explained here (as in \cite{M17}) that if one wishes to compare Riemann's $\xi$-function to Whittaker functions then their asymptotics for large negative $E$ dictate that $\kappa=\tfrac94$.
Just as for the Bessel function, however, the zeroes of the Whittaker function are too regularly spaced to match Riemann's $\xi$.  To avoid this, Jeffreys' condition $V'' \ll |V'|^{4/3}$ for his asymptotic treatment of turning points in the WKB method \cite{J}, must be broken.  If one wants to try to fit the eigenvalues of a potential to (the squares of) the zeroes of Riemann's $\xi$-function, as was attempted by \cite{WS}, though it seems they didn't realise that it is more sensible to use the squares, it is yet more sensible to try to fit the whole characteristic function to Riemann's $\xi$.  There is a proposed potential in \cite{Bo}, but its status is unclear to me.

\section{The decaying subspace}
In many numerical implementations of the shooting method, starting in the decaying subspace for $x \to +\infty$ is approximated by taking $\psi(b)=0, \psi'(b) = 1$ (or any unit vector in the $(\psi,\psi')$-plane, as in Weyl's theory \cite{T}) for some (or a sequence of) large $b$ and integrating leftwards from $b$.  

It is better, however, to start exactly in the decaying subspace, in particular if we want to study the dependence of the result on $E$, as here.  
This will be achieved under a WKB-like assumption on $V$ with $V(x) \to +\infty$ as $x \to +\infty$, by proving that for each $E \in \C$ there is a function on a halfline, representing the slope $S=\psi'/\psi$ of the decaying subspace for large enough $x$, and checking the solutions in it decay.  

Slopes $S$ of solutions to $L\psi = E\psi$ evolve by the Ricatti equation
\begin{equation}
S'=V-E-S^2.
\label{eq:Ric}
\end{equation}
They go through $S=\infty$ when $\psi$ goes through $0$ ($S(x)$ should really be considered as a point in $\C P^1$ and then there is no singularity on passing through $\infty$), but there are non-singular solutions if one restricts to large enough $x$, in particular so that $V(x)>\Re E$.

The decaying wavefunction corresponds to a solution near $S_0 = -\sqrt{V-E}$.  That approximation is a slow manifold for (\ref{eq:Ric}) in the sense that if $V$ and hence $S_0$ were constant then $S'$ would be zero and so it would be an exact solution, so if $V$ varies slowly, $S_0$ is nearly a solution.  For some theory and applications of slow manifolds, see \cite{M04}, and it is a subject on which Berry has written too.

For $V$ slowly varying, one can look for a solution as an asymptotic series, an idea that Berry has often used.  For example, one could generate it by iteration:
$$S_1 = -\sqrt{V-E-S_0'} = -\sqrt{V-E+\frac{V'}{2\sqrt{V-E}}},$$ and more generally
$$S_{n+1} = -\sqrt{V-E-S_n'}.$$  It is not clear, however, that it would converge.  In particular, it would depend on ever higher derivatives of $V$ so would probably require an analyticity assumption on $V$ to say anything much about it.  Indeed, the best one can achieve for many slow-manifold problems is a non-convergent asymptotic series.  

In ``normally hyperbolic'' cases, however, one can prove existence of an exact slow manifold, with as accurate bounds as desired, and this problem falls into that class.
This section will prove that there is an exact solution within a factor $\frac18$ of $S_0$, for $x$ large enough.

Given $E = E_r+ i E_i \in \C$, take $x_E$ sufficiently large that 
\begin{equation}
V(x)-E_r>\sqrt{2} |E_i|
\label{eq:cone}
\end{equation}
for all $x \ge x_E$ ($x_E$ may have to be increased a bit later, to satisfy another condition).  Then defining 
\begin{equation}
k = k_r + i k_i = \sqrt{V-E}
\label{eq:k}
\end{equation}
to be the branch with positive real part for all such $x$, it satisfies 
\begin{equation}
|k_i| < k_r.
\label{eq:ineq}
\end{equation}  
Under suitable conditions a non-singular solution $S$ of (\ref{eq:Ric}) will be obtained on $[x_E,\infty)$ that is relatively close to $-k(x)$, the approximate slow manifold for (\ref{eq:Ric}).  The solution $S$ repels other solutions as $x$ increases and corresponds to the slope of the decaying subspace.

Our solution is obtained as a fixed point of a map $C$ on a Banach space $\cC$ of $C^1$ functions:
\begin{equation}
C[S](x) = S(x) + \int_x^\infty e^{-2\int_x^y k(z)\, dz} (S'+S^2-k^2)(y)\ dy
\label{eq:defC}
\end{equation}
by iteration from initial guess $S = -k$.  The motivation is that this is the Newton step for finding a solution of the Ricatti equation (\ref{eq:Ric}), starting from the initial guess.
It will be shown to be a contraction on some ball around the initial guess if $V$ is differentiable with $|V'| \le 2c k_r^3$ on $[x_E,\infty)$ for some small enough $c$.
Our space $\cC$ of $C^1$ functions is those with $\|S\|$ finite, where
\begin{equation}
\|S\| = \max\left[\sup_{x\ge x_E} \frac{|S(x)|}{|k(x)|}, \sup_{x\ge x_E} \frac{|S'(x)|}{|k(x)|^2}\right].
\end{equation}
This norm is to allow for the expected growth of $S$ and $S'$.

First check that $C$ is defined, in particular that the outer integral in (\ref{eq:defC}) converges and the result is differentiable and in the space $\cC$.
The term $$|S'+S^2-k^2| \le (\|S\|+\|S\|^2+1) |k|^2.$$
Now $|k(y)|^2 \to +\infty$ as $y \to +\infty$, but for $x \ge x_E$, $|k|^2 < 2k_r^2$ by (\ref{eq:ineq}).  Let us assume that (possibly increasing $x_E$ if necessary)
\begin{equation}
|V' | \le 2c k_r^3
\label{eq:Vp}
\end{equation}
on $[x_E,\infty)$ for some $c<2$ (later $c$ will be taken to be $1/40$) 
(compare the WKB condition $|V'| \ll |k|^3$ and use $|k|^2 < 2k_r^2$ to see that this is satisfied by many potentials away from turning points; the WKB condition is usually applied in the oscillatory regime but can be used in the exponential regime too). 
From the definition (\ref{eq:k}), 
$$k' = \frac{V'}{2k}, \textrm{ thus } k_r' = \frac{V' k_r}{2|k|^2}.$$
It follows that 
\begin{equation}
|k_r' | \le c k_r^2,
\label{eq:krp}
\end{equation} 
and then $|e^{-2\int_x^y k(z)\, dz} |  = e^{-2\int_x^y k_r(z)\, dz}$ decays faster than $|k(y)|^2$ grows, thus the outer integral converges.  This is made explicit in the next paragraph.
From now on, shorten $\int_x^y k(z)\, dz$ to $\int_x^y k$ (and similarly for $\int k_r, \int S, \int T$) when they appear as the argument of an exponential.

To bound the outer integral in (\ref{eq:defC}), note that $$\frac{\partial }{\partial y} e^{-2\int_x^y k_r} = -2k_r(y) e^{-2\int_x^y k_r},$$ 
so integrating by parts yields
$$\int_x^\infty e^{-2\int_x^y k_r}\ 2k_r^2(y)\ dy = \int_x^\infty  e^{-2\int_x^y k_r} k_r'(y)\ dy +k_r(x).$$
Using $k_r' \le c k_r^2$ again, the integral on the right is at most $c/2$ times that on the left, and $c<2$ so 
\begin{equation}
\int_x^\infty e^{-2\int_x^y k_r}\ 2k_r^2(y)\ dy \le \frac{k_r(x)}{1-c/2}.
\label{eq:intbd}
\end{equation}
Hence 
\begin{equation}
|(C[S]-S)(x)| \le |k(x)|(\|S\|+\|S\|^2+1)/(1-c/2),
\label{eq:deltaC}
\end{equation}
in the form required.

The derivative of $C[S]$ is given by
\begin{eqnarray}
C[S]'(x) &=& S'(x) - (S'+S^2-k^2)(x) + \int_x^\infty 2k(x) e^{-2\int_x^y k}\ (S'+S^2-k^2)(y)\ dy \nonumber\\
&=& (k^2-S^2)(x) + 2k(x)(C[S]-S)(x).
\label{eq:deriv}
\end{eqnarray}
Using $|S|\le |k|\|S\|$ and the bound (\ref{eq:deltaC}), it follows that 
$$|C[S]'| \le |k|^2 \left(1+\|S\|^2 + 2\frac{\|S\|+\|S\|^2+1}{1-c/2}\right),$$ in the required form.


Secondly, bound the initial step
$\Delta = C[-k]+k$.
\begin{equation}
\Delta(x) = -\int_x^\infty e^{-2\int_x^y k}\, k'(y)\ dy.
\end{equation}
Using (\ref{eq:krp}),
$$|\Delta(x)| \le \int_x^\infty e^{-2\int_x^y k_r} c k_r^2(y)\, dy \le \frac{c/2}{1-c/2}k_r(x)$$
by (\ref{eq:intbd}).
So 
\begin{equation}
\frac{|\Delta|}{|k|} \le \frac{c/2}{1-c/2}.
\label{eq:step}
\end{equation}
Next, bound its derivative.
$$\Delta'(x) = k'(x) + \int_x^\infty 2k(x) e^{-2\int_x^y k}\, k'(y)\ dy = k'(x) - 2k(x)\Delta(x).$$
Thus
$$|\Delta'| \le ck_r^2+ 2|k|\frac{c/2}{1-c/2}k_r.$$
Hence
$$\frac{|\Delta'|}{|k|^2} \le c + \frac{c}{1-c/2} = \frac{2-c/2}{1-c/2}c.$$
Combining with (\ref{eq:step}) and using $c<2<3$, there follows
$$\|\Delta\| \le \sigma = \frac{2-c/2}{1-c/2}c.$$
It will be desirable shortly to reduce $c$ to make this small enough for a contraction map argument.


Thirdly, show that $C$ is a contraction for $S$ near $-k$.  This is easiest done by bounding its derivative $\delta C/\delta S$.  From (\ref{eq:defC}),
$$\delta C(x) = \delta S(x) + \int_x^\infty e^{-2\int_x^y k} (\delta S' + 2S\, \delta S)(y)\ dy.$$
Integrating by parts,
$$\int_x^\infty e^{-2\int_x^y k} \delta S' = \int 2k(y) e^{-2\int_x^y k}\delta S(y)\ dy -\delta S(x),$$
using  $|\delta S| \le |k| \|\delta S\|$ to show that the boundary term at infinity is zero.
Thus
$$\delta C(x) = \int_x^\infty e^{-2\int_x^y k}\ 2(k+S)\delta S(y) \ dy.$$
Suppose $\|S+k\| \le \eps$.  Then 
\begin{equation}
|S+k| \le \eps |k|.
\label{eq:spk}
\end{equation}
It follows that
$$|\delta C(x)| \le \int_x^\infty e^{-2\int_x^y k_r} 2\eps |k|^2 \|\delta S\|\ dy \le \int_x^\infty 4\eps k_r^2 e^{-2\int_x^y k_r}\, dy\ \|\delta S\| 
\le \frac{2\eps k_r(x)}{1-c/2} \|\delta S\|.$$
So
\begin{equation}
\frac{|\delta C|}{|k|} \le \frac{2\eps}{1-c/2}\|\delta S\|.
\label{eq:dC}
\end{equation}
Next, use (\ref{eq:deriv}) to obtain
$$\delta C' = -2S\,\delta S + 2k(\delta C-\delta S) = -2(k+S)\delta S +2k\,\delta C.$$
So
$$|\delta C'| \le 2\eps |k|^2 \|\delta S\| + 2|k|^2\tfrac{\eps}{1-c/2}\|\delta S\|.$$
It follows that
$$\frac{|\delta C'|}{|k|^2} \le 2\eps(1+\tfrac{1}{1-c/2})\|\delta S\|.$$
Combining this with (\ref{eq:dC}),
$$\|\delta C\| \le \alpha  \|\delta S\|,$$
where $$\alpha = 2\eps\tfrac{2-c/2}{1-c/2}.$$

Now, choose $c$ and $\eps$ to make $\sigma \le \eps (1-\alpha)$.  This guarantees that $C$ is a contraction map on the ball of radius $\eps$ around $-k$.  As a consequence there is a unique fixed point $S$ of $C$ in the ball.  
The inequality has a solution (an interval of $\eps$ around $\frac{1}{4}\frac{1-c/2}{2-c/2}$) iff 
$$c \le \tfrac{1}{8}(\tfrac{1-c/2}{2-c/2})^2.$$
For example, it suffices to choose $c = 1/40, \eps = 1/8$.
One could make larger $c$ work by increasing the $\sqrt{2}$ in (\ref{eq:cone}), which would make $k_r$ closer to $|k|$.

Lastly, having constructed the fixed point $S$, the decaying solutions are obtained from $\psi' = S\psi$, namely 
\begin{equation}
\psi(x) = \psi(x_E) e^{\int_{x_E}^x S}.
\label{eq:psi}
\end{equation}
They decay because $S$ has negative real part; more precisely, $S_r \le -\beta k_r$ with $\beta = 1-2\eps$, and $k_r(x) \to +\infty$ as $x \to \infty$.  The bound on $S_r$ comes from (\ref{eq:spk}) as follows:~$S_r+k_r \le \eps |k|$, so $S_r \le -k_r+\eps |k|$, but $|k|\le 2k_r$.

Note that the decay is fast enough to make $\psi_E$ square-integrable, because
$$\frac{|\psi(x)|^2}{|\psi(x_E)|^2} = e^{\int_{x_E}^x 2S_r} \le e^{-2\beta\int_{x_E}^x k_r}.$$
Write this as $2\beta k_r e^{-2\beta\int_{x_E}^x k_r} /2\beta k_r$ and integrate by parts to obtain
$$J = \int_{x_E}^\infty e^{-2\beta\int_{x_E}^x k_r}\ dx = \int_{x_E}^\infty e^{-2\beta\int_{x_E}^x k_r} \tfrac{k_r'}{2\beta k_r^2}\ dx +\frac{1}{2\beta k_r(x)}.$$
Now $|k_r'| \le c k_r^2$, so the integral on the right is at most $cJ/2\beta$.  Thus
$$J \le \frac{1}{(2\beta-c) k_r(x)},$$
hence
$$\int_{x_E}^\infty |\psi(x)|^2\ dx \le \frac{|\psi(x_E)|^2}{(2\beta-c)k_r(x)}.$$

Note that I think the claimed method in \cite{C} to produce a decaying solution from a positive solution on a halfline does not work.

\section{Asymptotics of decay}
Next it is shown that under a further condition on $V$, 
\begin{equation}
\int_{x_0}^\infty V(x)^{-1/2}\ dx < \infty,
\label{eq:sqrtV}
\end{equation}
for any two (non-zero) decaying solutions $\psi_E, \psi_F$ for $E,F \in \C$, $\lim_{x\to \infty} \tfrac{\psi_E(x)}{\psi_F(x)}$ exists.
It follows that one can normalise the decaying solutions for different $E$ to all be asymptotic to one of them, say $\psi_0$.

From (\ref{eq:psi}), 
$$\frac{\psi_E}{\psi_F}(x) = \frac{\psi_E}{\psi_F}(x_0) e^{\int_{x_0}^x (S_E-S_F)}.$$
So it is a question of showing that $\int_{x_0}^\infty S_E(y) - S_F(y)\, dy < \infty$.

Now $S_E(x)$ is complex differentiable with respect to $E$ for $x \ge x_E$.
Indeed, differentiating the equation $S' = V-E-S^2$ with respect to $E$, shows we should seek $\frac{\partial S}{\partial E}$ as a solution $\sigma$ of $\sigma' = -1 -2S\sigma$ going to zero at infinity, leading to
$$\sigma_E(x) = \frac{\partial S}{\partial E}(x) = \int_x^\infty e^{2\int_x^y S}\ dy,$$ which converges for $x \ge x_E$ because $\Re S$ is sufficiently negative ($|S+k| \le \eps |k|$, $\eps <1$ and $k_r > 0$ goes to infinity).
We can obtain an explicit bound by writing $$e^{2\int_x^y S} = (2S(y) e^{2\int_x^y S}) (2S(y))^{-1}$$ 
and integrating this by parts:
$$\sigma(x) = \int_x^\infty e^{2\int_x^y S} \tfrac{S'}{2S^2}(y)\ dy - \frac{1}{2S(x)}.$$
Then $S' = k^2-S^2 = (k+S)(k-S)$ and $|S+k|\le \eps |k|$ implies $|S'| \le \eps(2+\eps)|k|^2$.  Similarly, $|S|\ge(1-\eps)|k|$.
So $$|S'/S^2| \le \tfrac{2+\eps}{(1-\eps)^2}\eps .$$  It follows that
$$|\sigma| \le \frac{2+\eps}{2(1-\eps)^2}\eps |\sigma| + \frac{1}{2(1-\eps)|k|}.$$
Thus $$|\sigma| \le \frac{1}{2(1-\eps) (1- \frac{2+\eps}{2(1-\eps)^2}\eps)|k|} = \frac{1-\eps}{(2-6\eps+\eps^2)|k|}.$$

Then $$S_E(x) - S_F(x) = \int_F^E \sigma_G(x) \ dG$$ and
\begin{equation}
\left|\int_{x_0}^\infty \sigma_G(x) \ dx\right| \le \int_{x_0}^\infty  \frac{1-\eps}{(2-6\eps+\eps^2)|k|} \ dx <\infty
\label{eq:intsigma}
\end{equation}
if $\int_{x_0}^\infty V(x)^{-1/2}\ dx < \infty$.  Hence $\int_{x_0}^\infty S_E(x) - S_F(x)\, dx < \infty$.

For the chosen normalisation it results that
\begin{equation}
\psi_E(x_0) = \psi_0(x_0) e^{\int_{x_0}^\infty (S_0-S_E)}
\label{eq:psi0}
\end{equation}
for sufficiently large values of $x_0$ that $S_0$ and $S_E$ are non-singular on $[x_0,\infty)$.
Thus for $x \ge x_0$,
\begin{equation}
\psi_E(x) = \psi_0(x_0) e^{\int_{x_0}^x S_0 + \int_{x}^\infty (S_0- S_E)}.
\label{eq:psi2}
\end{equation}

\section{Construction of characteristic function}
In the case of boundary conditions $\psi(a)=0$, $\psi(x) \to 0$ as $x \to +\infty$, take the decaying solution $\psi_E(x)$ on $[x_E,\infty)$, normalised as above, continue it leftwards by $L\psi=E\psi$ to $x=a$ and let $$P_L(E) = \psi_E(a).$$  

The function $P_L$ is entire because it is complex differentiable for all $E \in \C$.  Indeed, one can compute $dP_L/dE$ by first differentiating (\ref{eq:psi0}):
$$\frac{\partial \psi_E}{\partial E}(x_0) = -\psi_E(x_0) \int_{x_0}^\infty \tfrac{\partial S}{\partial E}(x)\, dx.$$
This integral converges under the assumption of the previous section, as in (\ref{eq:intsigma}).
Next use $\psi' = S\psi$ to obtain $\frac{\partial \psi'}{\partial E}(x_0) = \frac{\partial S}{\partial E} \psi + S\frac{\partial \psi}{\partial E}$ at $x_0$.
Then propagate $(\psi_E,\psi'_E)$ and their $E$-derivatives from $x=x_0$ back to $x=a$ by $\psi'' = (V-E)\psi$, which has $E$-derivative $\frac{\partial \psi''}{\partial E} = -\psi + (V-E)\frac{\partial \psi}{\partial E}$, to obtain $\frac{\partial \psi}{\partial E}(a)$.  

The zeroes of $P_L$ are the eigenvalues of $L$, considered as an unbounded operator on $L^2([a,\infty),\C)$.  Firstly, if $P_L(E)=0$ then $\psi_E$ is a non-zero square-integrable solution of $L\psi=E\psi$ satisfying the boundary condition $\psi(a)=0$, so $E$ is an eigenvalue of $L$.  Secondly, if $E$ is an eigenvalue of $L$ then it has a square-integrable eigenvector $\psi$ with $\psi(a)=0$ and $L\psi = E\psi$; all solutions are a linear combination of the decaying solution $\psi_E$ and a growing one that is not square-integrable, thus the square-integrable ones are the multiples of $\psi_E$, so $P_L(E)=0$.

It is perhaps not important, but I believe $P_L$ has order in $[\tfrac12,1]$. Here is an incomplete argument.  By adding a constant to the potential, one can take $V>0$ everywhere and thus $x_0=a$.  For large $r$, I expect the maximum of $|P_L(E)|$ over $|E|=r$ to be taken at $E=-r$, but I don't have a proof of it.  Taking $E=-r$, then one can take $x_E = a$.  Write $T=S_0-S_E$.
From (\ref{eq:psi0}),
$$\psi_E(a) = \psi_0(a)e^{\int_{a}^\infty T},$$
and without loss of generality, take $\psi_0(a)=1$.
Now $S'=V-E-S^2$, so $$T' = E-S_0^2+S_E^2 = E-(S_0+S_E)T.$$
It follows that $$T(x) = -E \int_x^\infty e^{\int_x^y (S_0+S_E})\ dy.$$
So $$\int_a^\infty T(x)\ dx = r\int_a^\infty \int_x^\infty e^{\int_x^y (S_0+S_E)}\ dy \ dx = r\int_a^\infty \int_a^y e^{\int_x^y (S_0+S_E)}\ dx \ dy.$$
Now for $x \ll x_r$, defined to be the smallest such that $V(x_r) = r$, 
$$S_0+S_E \approx -V^{-1/2}-(V+r)^{-1/2} \approx - r^{-1/2}.$$  Thus
$e^{\int_x^y (S_0+S_E)} \approx e^{-r(y-x)}$ and $\int_a^y e^{\int_x^y (S_0+S_E)}\ dx \approx r^{-1/2}(1-e^{-r(y-a)})$.
Integrating this from $a$ to $x_r$ produces approximately $x_r r^{-1/2}$.  The rest of the range of integration produces something smaller.
So $\psi_E(a) \approx e^{x_r r^{1/2}}$.
Now $x_r \to \infty$ as $r \to \infty$ but can not be larger than $r^{1/2}$ else $\int_a^\infty V^{-1/2}(x)\, dx$ diverges.  Hence $P_L$ has order in $[\tfrac12,1]$.


In the case of boundary conditions $\psi(x) \to 0$ as $x \to \pm \infty$, the characteristic function is constructed by propagating normalised decaying solutions $\psi_E^\pm$ from each end to an arbitrary point $a$ and letting $$P_L(E) = ({\psi_E^+}' \psi_E^- - {\psi_E^-}' \psi_E^+)(a).$$ 
The result is independent of $a$, by conservation of the Wronskian.  The same conclusions that $P_L$ is entire and its zeroes are the eigenvalues follow.

One could extend the above construction to other cases of singular boundary value problems, for example, $V:(a,b) \to \R$ with $V(x) \to \infty$ as $x \to b$.  The conditions (\ref{eq:Vp}, \ref{eq:sqrtV}) on $V$ would need to be adapted.

\section{Examples}
If $V(x) = 4\pi^2 e^{2x}$ for $x>0$ with a hard wall at $x=0$, substitute $z = 2\pi e^x$ to see that (up to an arbitrary non-zero factor) $P_L(E) = K_{\sqrt{-E}}(2\pi)$, where $K_\nu$ is the modified Bessel function (alternatively known as MacDonald function; note that it is even in $\nu$ so the square root induces no singularity).  This is Polya's first approximation to $\xi(\omega)$, where $E=\omega^2/4$ \cite{P}.

If $V(x) = 4\pi^2 e^{4|x|}$ for $x \in \R$ then one obtains left and right-decaying solutions in terms of Bessel functions again (write $z = {\pi}e^{\pm 2x}$) and so $P_L(E) = K_{\tfrac12\sqrt{-E}}'({\pi})K_{\tfrac12\sqrt{-E}}({\pi})$.
Note that $K'_\nu(z) = -\tfrac12(K_{\nu-1}(z)+K_{\nu+1}(z))$, so we obtain an interesting variant of Polya's fake $\xi$-function 
$\xi^*(\omega) = K_{i\omega/2+9/4}(2\pi)+K_{i\omega/2-9/4}(2\pi)$ 
\cite{P}, corresponding to a product of the first approximation and an analogue of the second one.

If $V(x) = 8\pi^2 \cosh 4x$ for $x \in \R$ then one can express $P_L$ in terms of a modified Mathieu function.

If $V(x) = 4\pi^2e^{2x} -4\pi \kappa e^x$ on $x>0$ with hard wall at $x=0$ (a truncated Morse potential) then the substitution $z = 4\pi e^x$ shows that $P_L(E) = W_{\kappa,\sqrt{-E}}(4\pi)$, where $W$ is a Whittaker function (the case $\kappa=0$ is the Bessel function with argument multiplied by 2, up to a factor of square root of its argument).  

The potential $V$ can of course be scaled in magnitude and in $x$ and a version of $P_L$ scaled in $E$ is obtained.  The scalings above have been chosen so that the potentials have ``width function'' (the {\em width} $w(v)$ of a potential $V$ at height $v \in \R$ is the length of the set $\{x \in \R: V(x) \le v\}$)
$$w(v) = \log\frac{\sqrt{v}}{2\pi} + O(\frac{\log v}{\sqrt{v}}).$$  This ensures that the number of eigenvalues below a given energy $E$ agrees asymptotically with the number of Riemann zeroes below $2\sqrt{E}$ as $E \to \infty$.  Also addition of a constant $\gamma$ to $V$ is equivalent to subtracting $\gamma$ from $E$.
To compare these $P_L$ with Riemann's $\xi$-function, define $\Xi(E) = \xi(2\sqrt{E})$ and compare the logarithmic derivatives of $P_L$ and $\Xi$ (to remove the arbitrary factor in front of $P_L$).  

For the Whittaker, 
$$\frac{\partial }{\partial E} \log P_L(E)  \sim - \frac{1}{2\sqrt{-E}} \log \frac{\sqrt{-E}}{\pi} + \frac{\kappa-\tfrac12}{2E} + O((-E)^{-3/2})$$
as $E \to -\infty$.  The leading term agrees with that for $\Xi$ but agreement in the second term requires $\kappa = \tfrac94$.
One can also compare with the Whittaker function shifted by $\gamma$ in $E$ but it gives a third term $\frac{\gamma}{4}(-E)^{-3/2} \log\frac{\sqrt{-E}}{\pi e}$ that is intermediate in size between the second and the remainder, which does not occur for $\Xi$.  Thus one concludes that the best fit, as far as large negative $E$ asymptotics is concerned, is with $\kappa=\tfrac94$ and $\gamma=0$.  This tightens up the comparison made in \cite{L} (where even the prefactor $4\pi^2$ for the leading $e^{2x}$ term was not made explicit).
I call $\kappa$ in the Whittaker function the ``magnetic field'', because another place that Whittaker functions occur is as decaying solutions of the equation for eigenfunctions of the magnetic (or Maass) Laplacian with magnetic field $\kappa$ on the pseudosphere (of curvature $-1$) \cite{H}.

It would be interesting to try  fits to Riemann's $\xi$ by characteristic functions for some two-sided Schr\"odinger potentials. A two-sided version of the Whittaker potential is the Tzitzecka 
potential \cite{RS} $V(x)=4\pi^2 2^{-2/3} (2 e^{3x}+e^{-6x})$, where the scaling has been chosen to achieve the same width function; the substitution $z=e^{3x}$ produces a Whittaker-like equation, but perhaps this needs extending by a parameter like $\kappa$ to achieve a better fit.

As has already been remarked, potentials satisfying Jeffreys' condition $V'' \ll V'^{4/3}$ produce too regular oscillations in $P_L(E)$ for real positive $E$, so if one wishes to fit $\xi$ exactly then one would need to break this condition.

\section{Extensions}
The theory of 1D Schr\"odinger operators extends straightforwardly to Sturm-Liouville (SL) operators, $Ly = w^{-1}(-(py')' + qy)$ with $p$ and $w$ positive functions.  The above construction of a characteristic function extends under suitable conditions on $q,p,w$.  All SL operators can be transformed to Schr\"odinger form, however, so this is not a genuine extension.  Explicitly, consider SL operator on functions of $z$, let $Q = \log(pw)$ and change variables by $dx/dz = \sqrt{w(z)/p(z)}$ and $\psi = e^{Q/4} y$.  Then $e^{Q/4}Le^{-Q/4}$ is of Schr\"odinger form with potential $V(x) = \tfrac14 Q''(z)+\tfrac{1}{16}Q'(z)^2 + q(z)e^{-Q(z)/2}$.  On the other hand, for the eigenvalue problem $Ly = \lambda y$, the eigenvalue $\lambda$ has to be incorporated into $q$ as $\lambda w$, so it gives a generalised Schr\"odinger eigenvalue problem, which would require a modified treatment.

One could also extend the theory to multicomponent 1D SL operators.  Compare \cite{KK04} for the case of regular boundary conditions.  The way to extend the analysis of the present paper to this case is to recognise that the slope of the decaying solution generalises to the slope of the decaying subspace and that for SL problems that is a Lagrangian subspace so the slope is represented by a symmetric matrix-valued function $S$.  The point is that given $\psi$ in the decaying subspace at a point $x$, $\psi' = S\psi$ with $S$ symmetric.  $S$ satisfies the matrix version of the Ricatti equation and similar bounds could be obtained.

Lastly, the theory could be extended to Dirac operators in 1D.  A Dirac operator is $L$ of the form
$Ly = Jy' + P(x)y$
on even-dimensional vector functions $y$ with $P$ symmetric and $J=\left[\begin{array}{cc} 0 & I\\-I & 0 \end{array}\right]$.
For some background, see \cite{LeSa}.  This type of operator can be generalised by allowing $J$ to represent the inverse of more general symplectic forms.  Indeed, the eigenvalue problem for $L$ can be regarded as the equation for frequencies of a multisymplectic linear time-independent system in 1D space and time.  If the Hamiltonian for the time-evolution is positive-definite then the frequencies are all real and come in $\pm$ pairs \cite{AM,M86}.  This is an approach I suggested to Berry in 1997 as an alternative to seeking a Hermitian operator for the squares of the Riemann zeroes.  He rightly responded that it is equivalent, at least for standard Hamiltonians of the form kinetic-plus-potential with the canonical symplectic form. Yet in general, the relation is less evident.  I drafted a paper on this in January 2021, which perhaps I should develop into a publication some day.

\section*{Acknowledgements}
I am grateful to Jeffrey Lagarias for discussions on this topic during the MSRI programme on Hamiltonian dynamics, where we also explored de Branges canonical systems and spaces of entire functions (e.g.~\cite{R,ReS}), which are generalised eigenvalue versions of Dirac operators but I have not developed here.
Thus, part of the work was supported by the National Science Foundation under Grant No.~DMS-1440140 while the author was in residence at the Mathematical Sciences Research Institute in Berkeley, California, during the Fall 2018 semester.
But above all, I am grateful to Michael Berry for his endless stimulation and willingness to talk.

\end{document}